\documentclass[twocolumn,aps,prl,preprintnumbers,showpacs,nofootinbib]{revtex4-1}%%%Notitlepage, va todo junto%%%%
\usepackage[utf8]{inputenc}
\usepackage[english]{babel}
\usepackage{amsmath}
\usepackage{amsfonts}
\usepackage{amssymb}
\usepackage{epsfig}
\usepackage{graphics,psfrag,rotating}
\usepackage{graphicx}% Include figure files
\usepackage{dcolumn}% Align table columns on decimal point
\usepackage{bm}% bold math
\usepackage{epstopdf}

\newcommand{\lsim}   {\mathrel{\mathop{\kern 0pt \rlap
  {\raise.2ex\hbox{$<$}}}
  \lower.9ex\hbox{\kern-.190em $\sim$}}}
\newcommand{\gsim}   {\mathrel{\mathop{\kern 0pt \rlap
  {\raise.2ex\hbox{$>$}}}
  \lower.9ex\hbox{\kern-.190em $\sim$}}}
\begin{document}

\title{Focusing of geodesic congruences in an accelerated expanding Universe}
%\title{Accelerated expansion analysis from the Raychaudhuri equation}

%%%%%%%%%%%%%%%%%%%%%%%%%%%%%%%%%%%%%%%%%%%%%%%%
%%%%%%%%%%%%%%%%%%%%%%%%%%%%%%%%%%%%%%%%%%%%%%%%
%%%%%%%%%%%%%%%%%%%%%%%%%%%%%%%%%%%%%%%%%%%%%%%%

%\thispagestyle{empty}

\author{
F.\,D.\ Albareti$\,^{(a)}$\footnote{E-mail: fdalbareti@ucm.es},
J.\,A.\,R.\ Cembranos$\,^{(a)}$ \footnote{E-mail: cembra@fis.ucm.es} and
A.\ de la Cruz-Dombriz$\,^{(b,c)}$ \footnote{E-mail: alvaro.delacruz-dombriz@uct.ac.za}}

\affiliation{$^{(a)}$ Departamento de F\'{\i}sica Te\'orica I, Universidad Complutense de Madrid, E-28040 Madrid, Spain}
\affiliation{$^{(b)}$ Astrophysics, Cosmology and Gravity Centre (ACGC), University of Cape Town, 7701 Rondebosch, Cape Town, South Africa}
\affiliation{$^{(c)}$ Department of Mathematics and Applied Mathematics, University of Cape Town, 7701 Rondebosch, Cape Town, South Africa.}

\date{\today}

\pacs{ 04.50.-h, 98.80.-k, 04.20.-q}% PACS, the Physics and Astronomy

%04.20.-q	 Classical general relativity (see also 02.40.-k Geometry, differential geometry, and topology)
%04.20.Cv	Fundamental problems and general formalism
%04.50.-h	 Higher-dimensional gravity and other theories of gravity (see also 11.25.Mj Compactification and four-dimensional models, 11.25.Uv D branes)
%04.50.Kd	Modified theories of gravity
%98.80.-k	 Cosmology (see also section 04 General relativity and gravitation; for origin and evolution of galaxies, see 98.62.Ai; for elementary particle and nuclear processes, see 95.30.Cq; for dark matter, see 95.35.+d; for dark energy, see 95.36.+x; for superclusters and large-scale structure of the Universe, see 98.65.Dx)


%%%%%%%%%%%%%%%%%%%%%%%%%%%%%%%%%%%%%%%%%%%%%%%%%%%%%%%%%%%%%%%%%%%%%%%%%%%%%%%%%%%%%%%%%%%%%%%%%%%%%%%%%%%%%%%%%%%%%%%%%%%%%%%%%%%%%%%%%%%%%%%%%%%%%%%%%%%%%%%%%%%%%%%%
%Abstract
%%%%%%%%%%%%%%%%%%%%%%%%%%%%%%%%%%%%%%%%%%%%%%%%%%%%%%%%%%%%%%%%%%%%%%%%%%%%%%%%%%%%%%%
%%%%%%%%%%%%%%%%%%%%%%%%%%%%%%%%%%%%%%%%%%%%%%%%%%%%%%%%%%%%%%%%%%%%%%%%%%%%%%%%%%%%%%%

\begin{abstract}

We study the accelerated expansion of the Universe through its consequences on a congruence of geodesics. We make use of the Raychaudhuri equation which describes the evolution of the expansion rate for a congruence of timelike or null geodesics. In particular, we focus on the space-time geometry contribution to this equation. By straightforward calculation from the metric of a Robertson-Walker cosmological model, it follows that in an accelerated expanding Universe the space-time contribution to the Raychaudhuri equation is positive for the fundamental congruence, favoring a non-focusing of the congruence of geodesics. However, the accelerated expansion of the present Universe does not imply a tendency of the fundamental congruence to diverge. It is shown that this is in fact the case for certain congruences of timelike geodesics without vorticity. Therefore, the focusing of geodesics remains feasible in an accelerated expanding Universe. Furthermore, a negative contribution to the Raychaudhuri equation from space-time geometry which is usually interpreted as the manifestation of the attractive character of gravity is restored in an accelerated expanding Robertson-Walker space-time at high speeds.

\end{abstract}

\maketitle

%%%%%%%%%%%%%%%%%%%%%%%%%%%%%%%%%%%%%%%%%%%%%%%%%%%%%%%%%%%%%%%%%%%%%%%%%%%%%%%%%%%%%%%
%%%%%%%%%%%%%%%%%%%%%%%%%%%%%%%%%%%%%%%%%%%%%%%%%%%%%%%%%%%%%%%%%%%%%%%%%%%%%%%%%%%%%%%
%%%%%%%%%%%%%%%%%%%%%%%%%%%%%%%%%%%%%%%%%%%%%%%%%%%%%%%%%%%%%%%%%%%%%%%%%%%%%%%%%%%%%%%

\section{Introduction}

%%%%%%%%%%%%%%%%%%%%%%%%%%%%%%%%%%%%%%%%%%%%%%%%%%%%%%%%%%%%%%%%%%%%%%%%%%%%%%%%%%%%%%%
%%%%%%%%%%%%%%%%%%%%%%%%%%%%%%%%%%%%%%%%%%%%%%%%%%%%%%%%%%%%%%%%%%%%%%%%%%%%%%%%%%%%%%%
%%%%%%%%%%%%%%%%%%%%%%%%%%%%%%%%%%%%%%%%%%%%%%%%%%%%%%%%%%%%%%%%%%%%%%%%%%%%%%%%%%%%%%%
General Relativity (GR) as a theory that describes the geometrical structure of space-time and its connection with the energy content of the Universe has been highly successful in the last one hundred years. GR is not only able to give explanation of some discrepancies that had been arisen with Newtonian theory in the Solar System, but also provides a satisfactory description of the cosmological evolution of the space-time.

However, GR is not able to account for the latest cosmological and astrophysical observations with standard matter sources.
Among these observations are the anomalous rotational velocities of objects near the edge of the galaxies, the dynamics of galaxies in clusters, formation of large scale structures or gravitational lensing of background objects by galaxy clusters, such as the Bullet Cluster. For the explanation of these observations, more matter than the standard one is required, the so-called
dark matter (DM) which has an attractive gravitational character that decelerates the expansion of the Universe. Although there are
many plausible origins for this component \cite{DM}, DM is usually assumed to be in the form of thermal relics that
naturally freeze-out with the right abundance in many extensions of the standard model of particles \cite{WIMPs}.
Future experiments will be able to discriminate among the large number of candidates and models, such as direct
and indirect detection designed explicitly for their search \cite{isearches}, or even at high energy colliders, where they could be
produced \cite{Coll}. Furthermore, the authors of \cite{SIa}, recently awarded with the Nobel Prize in physics, discovered the accelerated expansion of the Universe. This result has been one of the main reasons for the revision of the cosmological evolution as predicted by GR. Neither the properties of usual matter and energy nor the believed properties of DM can account for this acceleration. One way of circumventing this problem is to postulating a new kind of fluid, the so-called dark energy, which has a non-standard equation of state allowing a positive energy density while having a negative pressure \cite{DE,cosmoproblema}. Another possibility to generate the present accelerated expansion of the Universe is by the modification of GR \cite{varios,fR}. Indeed, this type of modifications could produce an inflationary epoch in the very early Universe. This epoch is postulated to evade the problems with the initially homogeneous and isotropic state of the Universe which arise from considering particle horizons in Robertson-Walker cosmological models.

\smallskip
In this work,  we will analyze the accelerated expansion of the Universe by its description through its effects on a congruence of geodesics without assuming
any specific gravitational theory. This description has the advantage of being coordinate-independent. In an accelerated expansion it is expected that neighboring geodesics with similar direction will increasingly recede from each other. This evolution may be studied either through the geodesic deviation equation %MAY BE STUDIED THROUGH THE GEODESIC DEVIATION EQUATION 
\cite{Ellis-Elst} or alternatively by %OR ALTERNATIVELY BY (requires) 
the definition of a kinematic quantity, the expansion, which is quite analogous to the spatial divergence of a vector field. The expansion rate along the congruence will provide us with the appropriate tool to analyze the consequences of the cosmological accelerated expansion of the Universe. This rate is given by the Raychaudhuri equation (RE), which will be discussed and used throughout this work. RE was first derived by Raychaudhuri in \cite{Raychaudhuri} for congruences of timelike geodesics. Later, Sachs \cite{Sachs} introduced the corresponding RE for null geodesics. Ehlers \textit{et al.}\ \cite{Ehlers} provided a generalization of hydrodynamics to GR where results analogous to well-known theorems of hydrodynamics using RE equation were obtained. Given a congruence of timelike or null geodesics this equation provides the contraction or expansion rate of the congruence according to some kinematic properties of the congruence and the curvature of the space-time where the congruence is embedded. In GR, the RE together with assumed energy conditions guarantee the attractive character of gravity represented by a non-positive contribution to this equation from the space-time geometry. Furthermore, the RE has a fundamental role in the demonstration of important theorems due to Hawking and Penrose. These theorems assert that singularities of manifolds representing physical space-times are intrinsic to GR and not a special property of particular solutions of Einstein's equations provided that some energy conditions hold \cite{HE,Wald}.

\smallskip
In this investigation, we shall analyze the evolution of the expansion of a congruence of geodesics. This is carried out in a Robertson-Walker cosmological model from a kinematical point of view since no gravitational theory is assumed beyond the equivalence principle. This work is organized as follows: First, %since there is not unified convention for some quantities definitions in the gravitational theories context, we present our conventions and notation.
in Section \ref{raychaudhuriequation}, we present a short review of the derivation of the RE for illustrative purposes and stress some aspects usually disregarded in the available literature. Then, the kinematics of a congruence of geodesics in a Robertson-Walker space-time is studied in Section \ref{rwspace-times}.
We conclude the paper by giving our conclusions in Section \ref{conclusions}.
% Finally, the conclusions are presented.

%%%%%%%%%%%%%%%%%%%%%%%%%%%%%%%%%%%%%%%%%%%%%%%%%%%%%%%%%%%%%%%%%%%%%%%%%%%%%%%%%%%%%%%
%%%%%%%%%%%%%%%%%%%%%%%%%%%%%%%%%%%%%%%%%%%%%%%%%%%%%%%%%%%%%%%%%%%%%%%%%%%%%%%%%%%%%%%
%%%%%%%%%%%%%%%%%%%%%%%%%%%%%%%%%%%%%%%%%%%%%%%%%%%%%%%%%%%%%%%%%%%%%%%%%%%%%%%%%%%%%%%

% \section{Conventions and notation}
\section{Raychaudhuri equation derivation}
\label{raychaudhuriequation}

%%%%%%%%%%%%%%%%%%%%%%%%%%%%%%%%%%%%%%%%%%%%%%%%%%%%%%%%%%%%%%%%%%%%%%%%%%%%%%%%%%%%%%%
%%%%%%%%%%%%%%%%%%%%%%%%%%%%%%%%%%%%%%%%%%%%%%%%%%%%%%%%%%%%%%%%%%%%%%%%%%%%%%%%%%%%%%%
%%%%%%%%%%%%%%%%%%%%%%%%%%%%%%%%%%%%%%%%%%%%%%%%%%%%%%%%%%%%%%%%%%%%%%%%%%%%%%%%%%%%%%%
The  RE provides the expansion rate of a congruence of timelike or null geodesics. %It is advisable to discuss each case separately. Let us follow R.M.\ Wald \cite{Wald}
In the following, we will summarize the main results for both cases separately according to the notation in \cite{Wald}.
%derivation of Raychaudhuri equation that we summarize in this section.
For different reviews and approaches to the RE, we refer the reader to \cite{SayanKarandDadhich}.
%%%%%%%%%%%%%%%%%%%%%%%%%%%%%%%%%%%%%%%%%%%%%%%%%
In the following let us use abstract index notation (see \cite{Wald} for details). Thus, an expression like $\xi^a$ represents the vector itself and not a component in a certain basis. An expression like $g_{ab}\xi^a\phi^b$ must be understood as $\mathbf{g}\left(\boldsymbol{\xi},\boldsymbol{\phi}\right)$.
%  with the advantage that it distinguishes between a vector and its coordinates. 
When explicit mention to the components of a tensor in a certain basis is made, greek indices will be used, i.e.,\ $\xi^\mu$. %The principal advantage of this notation is that it differentiates equations that hold between tensors and equations that hold between their components in a certain basis.
The principal advantage of this notation is that it distinguishes equations that hold 
between tensors and 
equations that only hold for components in a certain basis.
%
%\smallskip
We use a metric signature ($-,+,+,+$) and
the  Riemann and Ricci tensors and the Ricci scalar are defined respectively as
$R_{abc}^{\;\;\;\;\;d}\equiv \partial_b \Gamma^d_{ac}-\partial_a \Gamma^d_{bc}+\Gamma^e_{ac}\Gamma^d_{eb}-\Gamma^e_{bc}\Gamma^d_{ea}\,$,
$R_{ac} \equiv R_{abc}^{\;\;\;\;\;b}$ and
$R\equiv R^{a}_{\ a}$.

\subsection{Congruences of timelike geodesics}

%%%%%%%%%%%%%%%%%%%%%%%%%%%%%%%%%%%%%%%%%%%%%%%%%
%%%%%%%%%%%%%%%%%%%%%%%%%%%%%%%%%%%%%%%%%%%%%%%%%

Let us consider a congruence of timelike geodesics and its tangent vector field $\xi^a$. It is assumed that the tangent vector field is normalized, i.e.,\ $\xi^a\xi_a=-1$, which is always feasible by choosing an affine parameterization of the congruence. %It is worth remembering that we have chosen a signature ($-,+,+,+$).
Let us define the (0,2) tensor
\begin{eqnarray}
\Xi_{ab} \equiv \nabla_b\xi_a\,.
\end{eqnarray}
It is easy to verify that
\begin{eqnarray}
\xi^a\Xi_{ab}=0\ \ \ \ \ ;\ \ \ \ \xi^b\Xi_{ab}=0\,.
\label{purespatial}
\end{eqnarray}
The first expression is a consequence of $\xi^a$ being normalized whereas the
second one is the geodesic equation. As usual the projector tensor $h_{ab}$ is defined by
\begin{eqnarray}
h_{ab} \equiv g_{ab}+\xi_a\xi_b\,.
\label{spatialmetric}
\end{eqnarray}
This is a spatial metric provided that $\xi^a$ is hypersurface orthogonal. This property of the vector field means that there exists a foliation of the manifold by hypersurfaces orthogonal to $\xi^a$. The metric induced by $g_{ab}$ on these hypersurfaces is $h_{ab}$.
For cases where %However, if $\xi^a$
an orthogonal hypersurface cannot be found, %hypersurfaces orthogonal to the vector field $\xi^a$ do not exist. Nevertheless,
the projector tensor $h_{ab}$ is still well-defined as the tensor that projects tensors in a subspace orthogonal to $\xi^a$ at each point of the manifold. It is worth noticing that $\Xi_{ab}$ has no components in the direction of $\xi^a$ according to expression %because of expression
\eqref{purespatial}. Consequently, $\Xi_{ab}$ is rather defined %in each point of the manifold
in the subspace orthogonal to $\xi^a$. This subspace is the hypersurfaces tangent space foliating the manifold when $\xi^a$ is hypersurface orthogonal. Finally, let us consider the following quantities
\begin{eqnarray}
\theta\equiv \Xi^{ab}h_{ab}\,,
\end{eqnarray}
\begin{eqnarray}
\sigma_{ab}\equiv \Xi_{\left(ab\right)}-\frac{1}{3}\,\theta h_{ab}\,,
\end{eqnarray}
\begin{eqnarray}
\omega_{ab}\equiv \Xi_{\left[ab\right]}\,;
\end{eqnarray}
which are the trace (with respect to the metric $h_{ab}$, but the trace is the same respect to $g_{ab}$ because of \eqref{purespatial}), the trace-free symmetric part and the antisymmetric part of $\Xi_{ab}$ respectively. These quantities are known as the expansion $\theta$, shear $\sigma_{ab}$ and twist $\omega_{ab}$ of the congruence. Thus, $\Xi_{ab}$ can be decomposed as
\begin{eqnarray}
\Xi_{ab}=\frac{1}{3}\,\theta h_{ab} +\sigma_{ab}+\omega_{ab}\,,
\end{eqnarray}
which is analogous to the decomposition of the deformation velocity tensor in hydrodynamics. The meaning of $\theta$, $\sigma_{ab}$ and $\omega_{ab}$ is clear from the analogy with hydrodynamics. The expansion $\theta$ takes into account the divergence or convergence of nearby geodesics. It may also be interpreted as a relative change in volume of a fluid particle. The shear $\sigma_{ab}$ provides information about the deformation of the geodesic congruence without change in volume. Finally, the twist $\omega_{ab}$ contains the vorticity of the congruence.
%
%\smallskip
%It is possible to derive
Equations for the rate of these kinematic quantities along each geodesic can be derived in the following way. After some calculus, it can be proved
\begin{eqnarray}
%\begin{aligned}
\xi^c\nabla_c\Xi_{ab}=%\xi^c\nabla_c\nabla_b\xi_a\\ &=\xi^c\nabla_b\nabla_c\xi_a+R_{cba}^{\;\;\;\;\;d}\xi^c\xi_d \\ &=\nabla_b\left(\xi^c\nabla_c\xi_a\right)-\left(\nabla_b\xi^c\right)\left(\nabla_c\xi_a\right)+R_{cba}^{\;\;\;\;\;d}\xi^c\xi_d \\
%&=
-\Xi^c_{\ b}\Xi_{ac}+R_{cba}^{\;\;\;\;\;d}\xi^c\xi_d\,,
%\end{aligned}
\label{larga}
\end{eqnarray}
where the Leibniz's rule and the geodesic equation satisfied by $\xi^a$ have been used.  Since we are interested in the equation for the expansion $\theta$,  the trace of \eqref{larga} becomes
\begin{eqnarray}
\frac{\text{d}\theta}{\text{d}\tau}=-\frac{1}{3}\,\theta^2-\sigma_{ab}\sigma^{ab}+\omega_{ab}\omega^{ab}-R_{ab}\xi^a\xi^b\,,
\label{ray}
\end{eqnarray}
where $\tau$ is an affine parameter of the geodesics, i.e., the proper time of an observer moving along a geodesic of the congruence. Equation \eqref{ray} is known as the RE \cite{Raychaudhuri, HE, Wald}. Let us recall that the expansion $\theta$ provides a natural coordinate-independent way of measuring the focusing or divergence of free-falling matter following a certain congruence of geodesics in the gravitational field, i.e., $\theta>0$ means divergence and $\theta<0$ means convergence. The quantities of the r.h.s.\ of \eqref{ray} are scalars, thus any convenient coordinate system can be used to calculate them. However, it is important to stress that equation \eqref{ray} provides the expansion rate as seen by an observer moving along a geodesic of the congruence.
The equivalent counterparts for shear $\sigma_{ab}$ and the twist $\omega_{ab}$ are obtained from
equation \eqref{larga} by taking the trace-free symmetric part and the antisymmetric part respectively.

\smallskip
Let us now discuss the character of the contribution of the different terms in expression \eqref{ray}. The quantities $\sigma_{ab}\sigma^{ab}$ and $\omega_{ab}\omega^{ab}$ are non-negative quadratic scalars. The fact that these scalars are non-negative is due to the {\it spatial} character of the tensor $\Xi_{ab}$ reflected by \eqref{purespatial}. This can be understood if  one considers a framework with
the spatial components of $h_{ab}$, i.e.,\ the orthogonal components to $\xi^a$, taking the values $\delta_{\mu\nu}$ with $\mu,\nu=1,2,3$ at a certain point of the manifold, whereas the components in the direction of $\xi^a$ being zero. Then, taking into account the spatial character of $\Xi_{ab}$ the values of $\sigma_{\mu\nu}$ and $\sigma^{\mu\nu}$ are the same in this special framework. Therefore, $\sigma_{ab}\sigma^{ab}\geq 0$. The same reasoning is valid for $\omega_{ab}$.

%
%%%%%% lo ha quitado Alvaro %%%%%%%%
%This can be understood if a coordinate system is used in which the spatial components of $h_{ab}$, i.e.,\ the orthogonal components to $\xi^a$, take the values $\delta_{\mu\nu}$ with $\mu,\nu=1,2,3$ in a certain point of the manifold while the components in the direction of $\xi^a$ being zero. This is possible because $h_{ab}$ has signature $(0,+,+,+)$. Thus
%%
%\begin{eqnarray}
%\sigma_{ab}\sigma^{ab}=g_{ac}\,g_{bd}\,\sigma^{cd} \sigma^{ab}=h_{ac}\,h_{bd}\,\sigma^{cd} \sigma^{ab}\,,
%\end{eqnarray}
%%
%where the spatial character of the tensor $\sigma_{ab}$ has been used. Then, in the special coordinate system described before the value of $\sigma_{\mu\nu}$ and $\sigma^{\mu\nu}$ is the same. Therefore, $\sigma_{ab}\sigma^{ab}\geq 0$. The same reasoning is valid for $\omega_{ab}$.
%%%%%%%%%%%%%%%%%
%
Thus, it is clear that both the shear $\sigma_{ab}$ and the expansion $\theta$ always tend the congruence to focus whereas the twist $\omega_{ab}$ always tends to make it diverge. For the contribution of space-time geometry, i.e., for the sign of $-R_{ab}\xi^a \xi^b$, there appear three possibilities:
\begin{eqnarray}
-R_{ab}\xi^a \xi^b > 0\ \ \ \ \ \ \text{Positive contribution} ,
\label{positive}
\end{eqnarray}
\begin{eqnarray}
-R_{ab}\xi^a \xi^b < 0\ \ \ \ \ \text{Negative contribution} ,
\label{negative}
\end{eqnarray}
\begin{eqnarray}
-R_{ab}\xi^a \xi^b = 0\ \ \ \ \ \ \ \ \ \ \text{Zero contribution}.
\label{zero}
\end{eqnarray}
The implications of the last three inequalities will be the keystone of this investigation. The effect that the two first possibilities produce in the congruence may be expressed as a tendency to diverge and a tendency to focus respectively. It can also be understood as accelerated expansion in the case of \eqref{positive} and decelerated expansion in the case of \eqref{negative}. On the other hand, the value of $R_{ab}\xi^a\xi^a$ depends both on the point of the manifold where it is calculated and on the direction of the vector field $\xi^a$ at that point. Thus, a positive contribution to the RE may be obtained at some points of the manifold for certain directions whereas may be negative at those points for different directions.

\smallskip
Although it is rarely considered, the expression $R_{ab}\xi^a\xi^b$ admits a geometrical interpretation \cite{Eisenhart}. Let $\xi^a_{(0)}$ be a unit vector and
$\xi^a_{(\alpha)}$ with $\alpha=1,2,3$ an arbitrary set of unit vectors orthogonal to $\xi^a_{(0)}$. The Gaussian curvature $K_{(0\alpha)}$ of the geodesic surface generated by $\xi^a_{(0)}$ and $\xi^a_{(\alpha)}$ is
\begin{eqnarray}
K_{(0\alpha)}=e_{(0)}\,e_{(\alpha)}\,R_{abcd}\,\xi^a_{(0)}\,\xi^b_{(\alpha)}\,\xi^c_{(0)}\,\xi^d_{(\alpha)}\,,
\end{eqnarray}
where $e_{(0)}$ and $e_{(\alpha)}$ are the usual Eisenhart $e$'s which take the values $+1$ or $-1$ according to $g_{ab}\,\xi^a_{(0)}\,\xi^b_{(0)}=e_{(0)}$ and $g_{ab}\,\xi^a_{(\alpha)}\,\xi^b_{(\alpha)}=e_{(\alpha)}$. %Special attention must be taken when dealing with Gaussian curvatures.
As is widely known, the sign convention for Riemann %$R_{ac}=R_{abc}^{\;\;\;\;\;b}$ tensor
and Ricci tensors %$R_{ab}$
is arbitrary a priori. However, let us recall that the sign of the Gaussian curvature possesses an intrinsic meaning sign-convention independent. % as it is well-known from classical differential geometry of surfaces.
Thus, when all these Gaussian curvatures $K_{(0\alpha)}$ are summed one gets %it is obtained
\begin{eqnarray}
%\begin{aligned}
\sum^{3}_{\alpha=0}K_{(0\alpha)}=%e_{(0)}\,R_{abcd}\,\xi^a_{(0)}\,\xi^c_{(0)}\,\sum^{3}_{\alpha=0}\,e_{(\alpha)}\,\xi^b_{(\alpha)}\,\xi^d_{(\alpha)}\\
%&=e_{(0)}\,R_{abcd}\,\xi^a_{(0)}\,\xi^c_{(0)}\, g^{bd}\\
%&=
e_{(0)}\,R_{ac}\,\xi^a_{(0)}\,\xi^c_{(0)},
%\end{aligned}
\label{sum}
\end{eqnarray}
where we have considered that $K_{(00)}$ is zero and the spectral decomposition of the metric
\begin{eqnarray}
g^{ab}=\sum^{3}_{\alpha=0}\,e_{(\alpha)}\,\xi^a_{(\alpha)}\,\xi^b_{(\alpha)}\,
\end{eqnarray}
has been used. Manifestly, the sum \eqref{sum} does not depend on the election of the set of orthogonal unit vectors $\xi^a_{(\alpha)}$ under consideration. Then, $e R_{ab}\xi^a\xi^b$ is the sum of all the Gaussian curvatures of the geodesics surfaces generated by $\xi^a$ and any set of orthogonal unit vectors to it. This has been referred to as the mean curvature in the direction of $\xi^a$% . For further details see
\cite{Eisenhart}.
Thus, by following our sign convention $e=-1$ and
the mean curvature ${\cal M}_{\xi^a}$ in the direction of a timelike vector $\xi^a$ becomes
\begin{eqnarray}
{\cal M}_{\xi^a}\,\equiv\, -R_{ab}\xi^a\xi^b\,.
\end{eqnarray}
Hence, the contribution of the space-time geometry to the RE \eqref{ray} has a clear geometrical interpretation as the mean curvature ${\cal M}_{\xi^a}$ in the direction of the tangent vector field $\xi^a$. Furthermore, the inequalities %three possibilities
\eqref{positive}, \eqref{negative} and \eqref{zero}
%the statement related to the contribution of space-time geometry to Raychaudhuri equation given in \eqref{positive}, \eqref{negative} and \eqref{zero}
can be rewritten in terms of the mean curvature ${\cal M}_{\xi^a}$ in the direction of $\xi^a$ as follows
\begin{eqnarray}
{\cal M}_{\xi^a}>0\ \ \ \ \ \ \text{Positive contribution} ,
\label{positiveM}
\end{eqnarray}
\begin{eqnarray}
{\cal M}_{\xi^a}<0\ \ \ \ \ \text{Negative contribution} ,
\label{negativeM}
\end{eqnarray}
\begin{eqnarray}
{\cal M}_{\xi^a}=0\ \ \ \ \ \ \ \ \ \:\ \text{Zero contribution}.
\label{zeroM}
\end{eqnarray}
Thus, if the mean curvature for a given timelike vector $\xi^a$ at one point of the manifold is positive (negative) the space-time geometry would tend to make geodesics with tangent vector $\xi^a$ passing through this point diverge (focus).
This formulation represents a clear advantage over the previous one in terms of $R_{ab}\xi^a\xi^a$ since it does not depend upon any sign convention nor upon the metric signature.
%%%%%%%%%%%%%%%%%%%
%Parrafo nuevo de Franco (16 agosto) que se une al anterior (Alvaro)
Let us recall that when energy conditions are assumed in GR, one always obtains \eqref{negativeM} and \eqref{zeroM} but not \eqref{positiveM}. The fact that in GR ${\cal M}_{\xi^a}$ is always negative or zero is interpreted as a manifestation of the attractive character of gravity.
%%%%%%%%%%%%%%%%%

\smallskip
Let us return to our discussion about the RE. It is clear that the sign of the expansion rate
$\text{d}\theta / \text{d}\tau$ will depend upon all the contributions of the r.h.s of \eqref{ray}. From their respectively definitions, it is obvious that all these terms depend on the congruence of geodesics under consideration.
However, the terms $\theta^2$, $\sigma_{ab}\sigma^{ab}$ and $\omega_{ab}\omega^{ab}$ depend on the behavior of the congruence in the neighborhood of the point where they are calculated. Let us recall that these terms come from $\nabla_{b}\xi_{a}$, then the tangent vector field $\xi^a$ must be known locally in order to calculate its covariant derivative. Nevertheless, the contribution of space-time geometry ${\cal M}_{\xi^a}$ depends only upon the direction of the congruence at a given point. Thus, all the congruences pointing in the same direction at a given point, will have the same contribution from space-time. %Therefore, the term ${\cal M}_{\xi^a}$ has more general implications.
Moreover, there are only two terms that may have a positive contribution: the term $\omega_{ab}\omega^{ab}$ and the one given by ${\cal M}_{\xi^a}$. Specifically, we shall restrict ourselves to congruences with $\omega_{ab}=0$ as we shall explain at the end of this section. Therefore, in this type of  congruences the only term that may have a positive contribution to \eqref{ray} is ${\cal M}_{\xi^a}$.

%%%%%%%%%%%%%%%%%%%%%%%%%%%%%%%%%%%%%%%%%%%%%%%%%
%%%%%%%%%%%%%%%%%%%%%%%%%%%%%%%%%%%%%%%%%%%%%%%%%

\subsection{Congruences of null geodesics}

%%%%%%%%%%%%%%%%%%%%%%%%%%%%%%%%%%%%%%%%%%%%%%%%%
%%%%%%%%%%%%%%%%%%%%%%%%%%%%%%%%%%%%%%%%%%%%%%%%%

%In order to get the expression \eqref{ray}, a congruence of timelike geodesics  was considered. If the interest one is interested in studying a congruence of null  geodesics, an
An equivalent version of the RE \eqref{ray} for null geodesics can be obtained by considering the null tangent vector field $k^a$ to a congruence of null geodesics \cite{HE, Wald, SayanKarandDadhich}. As expected the result is
\begin{eqnarray}
\frac{\text{d}\hat{\theta}}{\text{d}\lambda}=-\frac{1}{2}\,\hat{\theta}^2 -\hat{\sigma}_{ab}\hat{\sigma}^{ab}+\hat{\omega}_{ab}\hat{\omega}^{ab}-R_{ab} k^a k^b\,,
\label{raynull}
\end{eqnarray}
where the quantities $\hat{\sigma}_{ab}$, $\hat{\omega}_{ab}$ and $\hat{\theta}$ are defined with respect to the congruence of null geodesics, the tensor $\hat{h}_{ab}$ is the tensor analogous to $h_{ab}$ for null geodesics and $\lambda$ is an affine parameter in the following sense.
%An affine parameter of timelike geodesics is the one which normalizes the tangent vector field to the geodesics. Moreover, the geodesic equation when written in terms of an affine parameter takes the simple form of \eqref{geodesic}. If an arbitrary parameter is used, a term proportional to the tangent vector field will appear in the r.h.s.\ of \eqref{geodesic}.
When null geodesics are considered the tangent vector field has always zero norm and consequently an affine parameter defined as the one which normalizes the tangent vector field does not hold. For this reason, an affine parameter can be defined in a more general way as the one preserving the geodesic equation in its usual form. This definition remains valid both for timelike and null geodesics. The factor $1/2$ in \eqref{raynull} instead of $1/3$ appearing in \eqref{ray} is due to the fact that the involved subspace in the case of null geodesics is two-dimensional rather than three-dimensional. Thus, equation \eqref{raynull} is the RE for null geodesics.

\smallskip
%Let us recall that the contribution of the shear $\hat{\sigma}_{ab}$, twist $\hat{\omega}_{ab}$ and expansion $\hat{\theta}$ in the case of null geodesics has the same character than for timelike geodesics: expansion and shear tend to make the congruence of null geodesics focus while the twist tends to make it diverge.
%
Finally, let us remind that the same  type of  space-time geometry contributions provided by the inequalitites \eqref{positive}, \eqref{negative} and \eqref{zero} applies replacing $R_{ab}\xi^a\xi^b$ by $R_{ab} k^a k^b$. However, an analogous interpretation of the term $R_{ab} k^a k^b$ as a mean curvature is not practicable in this case \cite{Eisenhart}.

%%%%%%%%%%%%%%%%%%%%%%%%%%%%%%%%%%%%%%%%%%%%%%%%%
%%%%%%%%%%%%%%%%%%%%%%%%%%%%%%%%%%%%%%%%%%%%%%%%%
\subsection{Further remarks}
%%%%%%%%%%%%%%%%%%%%%%%%%%%%%%%%%%%%%%%%%%%%%%%%%
%%%%%%%%%%%%%%%%%%%%%%%%%%%%%%%%%%%%%%%%%%%%%%%%%

At this stage, we estimate important to stress that the Raychaudhuri equations \eqref{ray} and \eqref{raynull} are identities, i.e., geometrical statements which do not require any underlying gravitational theory to be derived. They are quite general and apply whenever a Riemannian\footnote{Let us stress that we use the adjective Riemannian in an extended sense, i.e.,\ Lorentzian metrics are included.} geometry
%Let us stress that we use the adjective Riemannian in an extended sense, i.e.,\ Lorentzian metrics are included.}
for the space-time is assumed.

\smallskip
A more general equation for the evolution of the expansion of a congruence results if the assumption of being a congruence of geodesics is discarded. Moreover, the RE can be generalized to spaces with torsion \cite{Capozziello}. For our purposes \eqref{ray} and \eqref{raynull} are sufficient.

\smallskip
As was briefly mentioned, there are analogous equations to \eqref{ray} and \eqref{raynull} for the shear $\sigma_{ab}$($\hat{\sigma}_{ab}$) and twist $\omega_{ab}$($\hat{\omega}_{ab}$) that provide the rate of these quantities along geodesics. These equations form a coupled system of non-linear first order differential equations. They also receive the generic name of Raychaudhuri equations. For instance, the equation for the twist $\omega_{ab}$ is an homogeneous differential equation in $\omega_{ab}$ and thus if $\omega_{ab}$ is zero initially it will remain zero along the geodesics. This result does not prove that the unique solution is $\omega_{ab}=0$ because of the non-linearity of the equation. However, this conclusion can be eventually proved more rigorously (see \cite{Ehlers}).  This is in fact the analogous result to Kelvin's theorem in hydrodynamics showing the analogy between  theorems of hydrodynamics and theorems concerning congruences of geodesics \cite{Ehlers}.

\smallskip
The initial value problem of Raychaudhuri equations is still an open question. However, in the case of equation \eqref{ray} if congruences that satisfy $\omega_{ab}=0$ are considered, there would exist hypersurfaces orthogonal to the vector field that foliate the space-time. This is a consequence of Frobenius' theorem \cite{Wald}.
Therefore, for a %arbitrary smooth
normalized vector field defined orthogonally to an arbitrary hypersurface -- not necessarily in the whole manifold --
its shear and its expansion (with components only in that hypersurface \eqref{purespatial}) can be determined. This way, the
expansion rate of the congruence of geodesics defined by this vector field can also be determined.
%, when evaluated in the orthogonal hypersurface where the vector field is defined, can also be determined.

\smallskip
Therefore, the rate of the kinematic quantities of the congruence ($\theta$, $\omega_{ab}$ and $\sigma_{ab}$ for timelike geodesics and $\hat{\theta}$, $\hat{\omega}_{ab}$ and $\hat{\sigma}_{ab}$ for null geodesics) is influenced by the geometry of space-time. In particular, in the case of expansion $\theta$ ($\hat{\theta}$), the term $-R_{ab}\xi^a\xi^b$ ($-R_{ab} k^a k^b$) is the contribution of space-time geometry for timelike geodesics (null geodesics). Let us recall that the contribution of space-time geometry has more general implications than the other contributions since the former depends solely on the direction of the tangent vector field at one point but not on the local behavior of the tangent vector field as $\theta$ ($\hat{\theta}$), $\sigma_{ab}$ ($\hat{\sigma}_{ab}$) and $\omega_{ab}$ ($\hat{\omega}_{ab}$). Thus, the contribution of space-time at one point of the manifold is the same for every congruence of geodesics with the same value of the tangent vector field at that point. However, the quantities $\theta$, $\sigma_{ab}$ and $\omega_{ab}$ will be different if the congruences are not locally the same.

\smallskip
Finally, let us remind here that the field equations derived from the assumed gravitational theory determine the geometry of the space-time.  For instance, in the case of GR, the usual Einstein's field equations enable to relate the quantities $R_{ab}\xi^a\xi^b$ and $R_{ab}k^ak^b$ which appear in \eqref{ray} and \eqref{raynull} respectively with the energy-momentum tensor $T_{ab}$ in a very simple way. Hence, the expansion rate is straightforwardly related
with the energy-momentum content.
As a result, general properties of the behavior of the geodesic congruence can be established provided that some energy conditions on the energy-momentum tensor are assumed \cite{HE, Wald}. More precisely, if the strong energy condition and the null energy condition are assumed (although the last one also follows by continuity if the strong energy condition holds) one reaches {\it the geodesic focusing theorem}, which assures the convergence until zero size of a congruence of geodesics without vorticity which is initially focusing. This result is used in the demonstration of the singularity theorems by Hawking and Penrose and it
has recently drawn wide attention due to its prominent role in the holographic principle \cite{Bousso}.

%%%%%%%%%%%%%%%%%%%%%%%%%%%%%%%%%%%%%%%%%%%%%%%%%%%%%%%%%%%%%%%%%%%%%%%%%%%%%%%%
%%%%%%%%%%%%%%%%%%%%%%%%%%%%%%%%%%%%%%%%%%%%%%%%%%%%%%%%%%%%%%%%%%%%%%%%%%%%%%%%
%%%%%%%%%%%%%%%%%%%%%%%%%%%%%%%%%%%%%%%%%%%%%%%%%%%%%%%%%%%%%%%%%%%%%%%%%%%%%%%%

\section{Kinematic considerations in Robertson-Walker space-times}
\label{rwspace-times}

%%%%%%%%%%%%%%%%%%%%%%%%%%%%%%%%%%%%%%%%%%%%%%%%%%%%%%%%%%%%%%%%%%%%%%%%%%%%%%%%
%%%%%%%%%%%%%%%%%%%%%%%%%%%%%%%%%%%%%%%%%%%%%%%%%%%%%%%%%%%%%%%%%%%%%%%%%%%%%%%%

% The only assumption is that space-time may be represented mathematically by a four-dimensional manifold with a Riemannian metric defined in it. Moreover, the conclusions of this section will be generally valid provided that homogeneity and locally isotropy are assumed in flat spacelike hypersurfaces that foliate the space-time, i.e., a spatially flat Robertson-Walker space-time.

In this section, let us start by considering a spatially flat Robertson-Walker metric \cite{FLRW},
\begin{eqnarray}
\text{d}s^2=-\text{d}t^2+a^2(t)\left(\text{d}x^2+\text{d}y^2+\text{d}z^2\right)\,,
\label{FLRW}
\end{eqnarray}
that
%where $k$ holds for the Gaussian curvature of $t=\text{constant}$ hypersurfaces being
%$k=1,0$ or $-1$ its possible values.
%
%\smallskip
according to the ultimate measurements of cosmological densities of the Universe \cite{WMAP7},
represents correctly the Universe at sufficiently large scales in agreement with the hypotheses of homogeneity and isotropy stated in
the Cosmological principle.
\smallskip
%From now on, the time dependence in $a(t)$ will be omitted when it is clear from context.
%
%the Universe is approximately spatially flat $k \simeq 0$. In consequence, we will take $k=0$ and cartesian coordinates for the space will be used. Thus the metric \eqref{FLRW} becomes
%
%\begin{eqnarray}
%\text{d}s^2=-\text{d}t^2+a^2(t)\left(\text{d}x^2+\text{d}y^2+\text{d}z^2\right)\,.
%\label{flat}
%\end{eqnarray}
%
%%%%%%%%%%%%%%%%%%%%%%%%%%%%%%%%%%%%%%%%%%%%%%%%%%%%%%%%%%%%
%For reference, let us sketch here some calculations. The non-vanishing Christoffel symbols for the Levi-Civita connection derived from \eqref{flat} are
%
%\begin{eqnarray}
%\Gamma^t_{xx}=\Gamma^t_{yy}=\Gamma^t_{zz}=a\dot{a}
%\end{eqnarray}
%
%\begin{eqnarray}
%\Gamma^x_{xt}=\Gamma^x_{tx}=\Gamma^y_{yt}=\Gamma^y_{yt}=\Gamma^z_{zt}=\Gamma^x_{tz}=\frac{\dot{a}}{a}\,,
%\end{eqnarray}
%
%where dot holds for the cosmic time $t$ derivative. %{a}=\frac{da}{dt}$.
%The Ricci tensor components are
%
%\begin{eqnarray}
%R_{tt}=-3\frac{\ddot{a}}{a}\ \ ,\ \ R_{xx}=R_{yy}=R_{zz}=a \ddot{a}+2\dot{a}^2\,;
%\label{Ricciterms}
%\end{eqnarray}
%
%the rest of them being zero. Finally, the Ricci scalar is
%
%\begin{eqnarray}
%R=6\left(\frac{\ddot{a}}{a}+\frac{\dot{a}^2}{a^2}\right)\,.
%\end{eqnarray}
%
The connection between the inequality \eqref{positive} -- or equivalently \eqref{positiveM} -- with the accelerated expansion of the Universe is not straightforward. The accelerated expansion of the Universe in a Robertson-Walker cosmological model is governed by the condition $\ddot{a}>0$  where the derivatives are taken with respect to cosmic time.
However, provided that the inequality \eqref{positiveM} holds for all the timelike directions $\xi^a$
 in a region of the manifold, the contribution of space-time geometry then favors
 a tendency to diverge of all the congruence of geodesics in that region.
Therefore, the natural question that arises is whether the condition $\ddot{a}>0$, when satisfied in a region, implies ${\cal M}_{\xi^a}>0$ for every timelike direction $\xi^a$. This question will be addressed throughout the present section. % in the region where $\ddot{a}>0$ holds We shall shown that this is not the case.

\smallskip
One may thus wonder about the effects of the cosmological acceleration in the kinematical behavior of a congruence of timelike geodesics.
%The accelerated expansion of the Universe may be reflected in the kinematical behavior of a congruence of timelike geodesics.
In particular, it is of special interest to study whether a congruence of timelike geodesics tends to focus -- or to diverge -- and under
which conditions this happens. The kinematic quantity of a congruence known as expansion $\theta$ introduced in Section \ref{raychaudhuriequation} provides us with a useful tool to study the focusing of timelike geodesics.
In this sense, the inequality %holds
\begin{eqnarray}
\xi^c\nabla_c\theta\,=\,\frac{\text{d}\theta}{\text{d}\tau} > 0\,,
\label{expansion}
\end{eqnarray}
states the tendency of nearby geodesics to diverge measured by a comoving observer. The proper time $\tau$ is the time of the comoving observer.
Since the l.h.s.\ of the previous expression \eqref{expansion}
can be replaced by the r.h.s.\ of the RE \eqref{ray},
a vector field tangent to a congruence of timelike geodesics must be chosen in order to determine
the expansion $\theta$, shear $\sigma_{ab}$ and twist $\omega_{ab}$.
% in order to know all the terms that appear in the r.h.s.\ of expression \eqref{ray}.

%%%%%%%%%%%%%%%%%%%%%%%%%%%%%%%%%%%%%%%
%%%%%%%%%%%%%%%%%%%%%%%%%%%%%%%%%%%%%%%

\subsection{Fundamental congruence}
\label{fundamentalcongruence}

%%%%%%%%%%%%%%%%%%%%%%%%%%%%%%%%%%%%%%%
%%%%%%%%%%%%%%%%%%%%%%%%%%%%%%%%%%%%%%%

Since we are dealing with a Robertson-Walker metric \eqref{FLRW}, the most natural choice for a vector field would be
\begin{eqnarray}
\xi^a_{\text{RW}}=\frac{\partial}{\partial t}\,,
\label{RW}
\end{eqnarray}
which is the preferred direction of a Robertson-Walker space-time because of the local isotropy in hypersurfaces orthogonal to $\xi^a_{\text{RW}}$. The vector field \eqref{RW} gives rise to a congruence of timelike geodesics of the form % as one can easily checked from expression \eqref{geodesic}.
%
%The congruence of geodesics  defined by \eqref{RW}
%are
$x^\mu=\left(t-t_{0},\vec{x}_{0}\right)$ in comoving coordinates with $t_{0}$ and $\vec{x}_{0}$ constants
\footnote{Therefore, being {\it steady} is a geodesic motion in Robertson-Walker metric.}.
%\footnote{Therefore, being ``steady'' is a geodesic motion in Robertson-Walker metric.}.
These geodesics are followed by the fundamental observers, those seeing homogeneity and isotropy in constant-time hypersurfaces. Thus, a fundamental congruence refers to the congruence of timelike geodesics generated by $\xi^a_{\text{RW}}$. % (see %. An interesting discussion of the kinematic quantities of the fundamental congruence of a Robertson-Walker space-time can be found in
%\cite{PaulUllrich} for further details).
As it is easy to verify by straightforward calculation -- or by using symmetry arguments --, the kinematic quantities of the fundamental congruence are all zero except the expansion
\begin{eqnarray}
\theta_{\text{RW}}=3\frac{\dot{a}}{a}\,.
\label{expansionRW}
\end{eqnarray}
Thus, by inspection of \eqref{ray} it follows that the rate of the expansion seen by a fundamental observer will be entirely determined by the value of the expansion $\theta_{\text{RW}}$ and the contribution of space-time geometry
\begin{eqnarray}
{\cal M}_{\xi^a_{\text{RW}}}=-R_{ab}\xi^a_{\text{RW}}\xi^b_{\text{RW}}=3\frac{\ddot{a}}{a}\,.
\label{fundamentalmean}
\end{eqnarray}
Hence
\begin{eqnarray}
\frac{\text{d}\theta_{\text{RW}}}{\text{d}t}\,=\,%-\frac{1}{3}\left(3\frac{\dot{a}}{a}\right)^2+3\frac{\ddot{a}}{a}  \,=\,
3\left[-\left(\frac{\dot{a}}{a}\right)^2+\frac{\ddot{a}}{a}\right],
\label{dtexpansionRW}
\end{eqnarray}
where the proper time of the fundamental observer is $t$, i.e.,\ the time coordinate in \eqref{FLRW}.
%Equation \eqref{dtexpansionRW} also follows by differentiation of  \eqref{expansionRW} with respect to coordinate $t$.
%%%%% he quitado las siguientes tres l?neas porque no dicen nada nuevo, solo dan un consejo.
%These manipulations may seem trivial but one should not forget
%that the mean curvature in the direction $\xi^a$, i.e.,\ ${\cal M}_{\xi^a}=-R_{ab}\xi^a\xi^b$, may be related to the energy-momentum tensor $T_{ab}$ through the field equations provided by the corresponding gravitational theory. Therefore, it is important to know at which step the term $-R_{ab}\xi^a\xi^b$ appears.
%
A fundamental observer will then see nearby geodesics with a tendency to diverge if and only if $\text{d}\theta_{\text{RW}}/\text{d}t>0$.
By imposing this condition on \eqref{dtexpansionRW} we get
\begin{eqnarray}
\ddot{a}>\frac{\dot{a}^2}{a}\,,
\end{eqnarray}
which can be expressed in the following way
%
%\begin{eqnarray}
%q<-1\ \ \ \ \ \Rightarrow \ \ \ \ \ \frac{\text{d}\theta_{RW}}{\text{d}t}>0\,,
%\label{deceleratingparameter}
%\end{eqnarray}
%
%
\begin{eqnarray}
q\equiv-\frac{1}{H^2}\frac{\ddot{a}}{a}\,<\,-1 \ \ \ \ \ \Leftrightarrow \ \ \ \ \ \frac{\text{d}\theta_{RW}}{\text{d}t}>0\,, %\ \ \ \text{and} \ \ \ H\equiv\frac{\dot{a}}{a}\,
\label{deceleratingparameter}
\end{eqnarray}
where $q$ is the deceleration parameter and $H\equiv \dot{a}/a$ is the Hubble parameter. Hence, a fundamental observer will perceive nearby geodesics with a tendency to diverge if and only if $q<-1$. In other words, getting a deceleration parameter smaller than zero is not sufficient for fundamental timelike geodesics to be perceived  by a fundamental observer as diverging in an increasingly fashion.

\smallskip
In the standard cosmological  model $\Lambda\text{CDM}$, the value of $q$ today can be determined by using the values of the present relative abundances
of matter (including dark matter) and cosmological constant, $\Omega_{M}$ and $\Omega_{\Lambda}$ respectively by the expression
\begin{eqnarray}
q=\frac{1}{2}\Omega_{M}-\Omega_{\Lambda}\,,
\end{eqnarray}
where flatness in spatial sections was again assumed. % Since we are not interested in the exact value of $q$, approximate values will be used.
Thus, using the approximate values $\Omega_{M} \simeq 0.27$ and $\Omega_{\Lambda} \simeq 0.73$ given in \cite{WMAP7}, the present value of the deceleration parameter $q_{0}$ results
%
%\begin{eqnarray}
$q_{0} \simeq -0.6$,
%\label{qvalue}
%\end{eqnarray}
%
with an uncertainty of the order of $0.01$. This value of $q_{0}$ does not imply a tendency of nearby geodesics to diverge since $q_{0}>-1$. One may argue that we have assumed GR to calculate the deceleration parameter $q_{0}$, and therefore the discussion is not independent of a gravitational theory. However, there are straightforward estimations of the deceleration parameter using Supernovae Ia \cite{Lazkoz} without appealing to a gravitational theory. The preferred values obtained for $q_{0}$ in these analyses are always larger than $-1$, although with uncertainties of the order $0.1$.

\smallskip
Before continuing with the study of the kinematics of a congruence of timelike geodesics, let us mention that
in general, the determination of other geodesics  is not possible in a closed form since an expression for $a(t)$ is usually missing.
%Even in these cases, the geodesics may be determined by numerical methods.
%
%
Therefore, it will not be possible in general to obtain an expression for the tangent vector field $\xi^a$ of a congruence of geodesics. However, if our interest is focused on calculating the space-time geometry contribution to the RE for timelike geodesics,
%, i.e., determining ${\cal M}_{\xi^a}$, at one point of the manifold,
the knowledge of the vector field $\xi^{a}$ expression is not necessary. The reason is that the value of ${\cal M}_{\xi^a}$ only depends on the direction of the vector field at a given point. Then, given a timelike vector, the value of ${\cal M}_{\xi^a}$ can be determined by using the expressions for the components of the Ricci tensor. Moreover, the value of ${\cal M}_{\xi^a}$ will be constant in each hypersurface orthogonal to $\xi^a_{\text{RW}}$ because of homogeneity in these hypersurfaces.

\smallskip
However, if we wish to know the total value of $\text{d}\theta/\text{d}\tau$ from the RE, an expression for the tangent vector field is needed. It means that we must know the vector field locally at each point of the manifold. Nevertheless, we can use the property stated in equations \eqref{purespatial} which asserts the {\it spatial} character of $\Xi_{ab}=\nabla_{b}\xi_a$. The procedure consists on defining a smooth vector field on a spacelike hypersurface. The only conditions on the definition of this vector field are that it be normalized $\xi^a\xi_a=-1$ and that it be orthogonal to the hypersurface where it is defined. This slice of a vector field will give rise to a congruence of timelike geodesics passing by the hypersurface with the direction of the vector field in each point. This congruence will have $\omega_{ab}=0$ because being zero in the initial hypersurface it will remain so. However, the shear $\sigma_{ab}$ and the expansion $\theta$ do not need to be zero. Then, instead of considering a vector field tangent to a congruence of geodesics defined in all the manifold, a vector field is instead defined at a spacelike hypersurface. Let us remark that this procedure restricts us to consider congruences of timelike geodesics without vorticity. By this way $\Xi_{ab}$ at the initial hypersurface can be determined. Thus, the expansion rate $\text{d}\theta/\text{d}\tau$ of the congruence evaluated in the hypersurface will be known from a slice of the geodesic tangent vector field.

%%%%%%%%%%%%%%%%%%%%%%%%%%%%%%%%%%%%%%%
%%%%%%%%%%%%%%%%%%%%%%%%%%%%%%%%%%%%%%%

\subsection{One-parameter dependent congruences of timelike geodesics}

%%%%%%%%%%%%%%%%%%%%%%%%%%%%%%%%%%%%%%%
%%%%%%%%%%%%%%%%%%%%%%%%%%%%%%%%%%%%%%%

The congruences under consideration from now on are obtained from a one-parameter dependent slice of a vector field. First of all, we must choose a spacelike hypersurface where the vector field will be defined. The simplest one-parameter dependent $\epsilon$ family of spacelike hypersurfaces defined by a function $f=0$ in terms of the coordinates \eqref{FLRW} is
\begin{eqnarray}
f=t-\vec{\epsilon} \cdot \vec{x}=0\,,
\label{hypersurfaces}
\end{eqnarray}
where $0 \leq \lvert\vec{\epsilon}\,\rvert/a < 1$ in order the hypersurfaces to be spacelike. This constraint is easily understood if one considers the normal vector to the hypersurfaces \eqref{hypersurfaces} and this vector is then constrained to be timelike. This family of hypersurfaces depends in fact on three parameters, namely the three components of $\vec{\epsilon}$. However, thanks to the spatial isotropy, the direction of $\vec{\epsilon}$ does not affect the results. Thus, the relevant parameter is $\lvert\vec{\epsilon}\,\rvert=\epsilon$. Moreover, because of spatial homogeneity and the arbitrariness in the election of the time origin we have eliminated the constant that may appear in \eqref{hypersurfaces}, i.e.,\ the hypersurfaces pass through the origin of coordinates. The above restriction on the value of $\epsilon$ means that the hypersurface is defined only in the manifold region $a > \epsilon$. Otherwise, the hypersurfaces are not spacelike. Let us now define the following quantity
\begin{eqnarray}
\beta\equiv \frac{\epsilon}{a^2(t)}\,.
\label{coordinatevelocity}
\end{eqnarray}
The slice of the timelike vector field which is normalized and orthogonal to the hypersurface given by \eqref{hypersurfaces} 
then becomes
\begin{eqnarray}
\xi^a_{(\beta)}=\tilde{\gamma}\left(\frac{\partial}{\partial t}+\beta \frac{\partial}{\partial x}\right)\,,
\label{generaltimelikevector}
\end{eqnarray}
where
\begin{eqnarray}
\tilde{\gamma} \equiv \frac{1}{\sqrt{1-a^2\beta^2}}.
\label{tilde}
\end{eqnarray}
The tangent vector field that gives rise to the fundamental congruence is obtained in the specific case $\epsilon=0$ or equivalently $\beta=0$.
In order to guarantee $\xi^a_{(\beta)}$ to be timelike, one imposes $0\leq a\beta < 1$. % in order to , i.e.,\ $g_{ab}\xi^a_{(\beta)}\xi^b_{(\beta)} < 0$, where $g_{ab}$ is the metric tensor as given in expression \eqref{flat}.
Then, it follows the restriction on the value of $\epsilon$ commented above. From \eqref{generaltimelikevector}, $\beta$ can be interpreted as the spatial coordinate velocity of a particle passing through the hypersurfaces defined by \eqref{hypersurfaces}. The dependence of $\beta$ on the scale factor $\eqref{coordinatevelocity}$ means that this velocity is not the same in the whole hypersurface defined by a certain value of the parameter $\epsilon$. Note that in expression  \eqref{generaltimelikevector} we have chosen the $x$-axis in the direction of the spatial coordinate velocity $\beta$ without loss of generality thanks to the spatial isotropy. The $\tilde{\gamma}$ factor just ensures the normalization $g_{ab}\xi^a_{(\beta)}\xi^b_{(\beta)}=-1$.

One realizes of the similarity of the former derivation
%factor $\tilde{\gamma}$
with the usual definitions of $\beta$  and $\gamma$ in Special Relativity. Consequently, % $\gamma=1/\sqrt{1-\beta^2}$. In Special Relativity, the parameter $\beta$ is the velocity in non-dimensional units, i.e.,\ $\beta=v/c$. The maximum velocity that a physical entity could reach is $c$, consequently $0\leq \beta <1$. In other words,
the slice of the timelike vector field orthogonal to the hypersurfaces \eqref{hypersurfaces} in Special Relativity becomes
\begin{eqnarray}
u^a=\gamma\left(\frac{\partial}{\partial t}+\beta \frac{\partial}{\partial x}\right)
\end{eqnarray}
with $0\leq \beta<1$ in order to guarantee $\eta_{ab} u^a u^b <0$, with $\eta_{ab}$ the Minkowskian metric. %, in $c=1$ units,
%s
%\begin{eqnarray}
%\text{d}s^2=-\text{d}t^2+\left(\text{d}x^2+\text{d}y^2+\text{d}z^2\right)\,.
%\end{eqnarray}
%
%It follows that in Minkowski space-time the scale factor is $a=1$ obviously constant.
Hence, expressions \eqref{generaltimelikevector} and \eqref{tilde} are completely valid both in Special Relativity (setting $a=1$) and in Robertson-Walker space-times. % The speed of light $c$ provides us with a conversion factor between spatial units and cosmological time.
However, when we are considering a Robertson-Walker metric, the maximum coordinate velocity $\beta_{max}$ that a physical entity can reach, in the sense that it moves along null geodesics, depends on the value of scale factor $a(t)$ at each point.
In order to guarantee that the vector field as given by \eqref{generaltimelikevector} is timelike, the required condition reads $0\leq a\beta <1$. Therefore the maximum coordinate velocity is expressed as $\beta_{max}=1/a(t)$.
%%%%%%% A?adido de Franco (16 agosto)
Nevertheless, let us recall that the physical spatial velocity which is measured is $a\beta$.
One may interpret a physical velocity as the ratio between a {\it spatial} and {\it time} distances. These distances hold for the ones 
induced by the timelike geodesic one follows.
Therefore, for a particle with four-velocity $\xi^a_{(\beta)}$, its physical velocity yields
\begin{eqnarray}
{\cal V}_{physical}\,=\,\lim_{\Delta t \rightarrow 0}\frac{\int_{t}^{t+\Delta t} {\sqrt{h^{{\rm RW}}_{ab}\xi^a_{(\beta)} \xi^b_{(\beta)}}\  \text{d}t'}}{\int_{t}^{t+\Delta t}{-\xi^{\rm RW}_{a}\xi^a_{(\beta)}\ \text{d}t'}}
% \nonumber \\
% \stackrel{\Delta t \rightarrow 0 }{\longrightarrow} \frac{\sqrt{h^{RW}_{ab}\xi^a_{(\beta)} \xi^b_{(\beta)}}\  \Delta t}{-\xi^{RW}_{a}\xi^a_{(\beta)}\ \Delta t}=
= a\beta
\label{physicalvelocity}
\end{eqnarray}
where $\xi^{\rm RW}_{a}$ is given by \eqref{expansionRW}
and $h^{\rm RW}_{ab}=g_{ab}+\xi^{\rm RW}_a\xi^{\rm RW}_b$ is the spatial metric induced by the flow of time in the direction of $\xi^a_{\rm RW}$. Alternatively, one may also build a unit vector orthogonal to $\xi^a_{\rm RW}$ being the spatial projection of $\xi^a_{(\beta)}$,
namely $h_{a}\equiv h^{\rm RW}_{ab} \xi^b_{(\beta)}/\sqrt{g^{ac} h^{\rm RW}_{ab} \xi^b_{(\beta)} h^{\rm RW}_{cd} \xi^d_{(\beta)}}$,
and then calculate the ratio between the projection of $\xi^a_{(\beta)}$ in the direction of $h_{a}$ and in the direction of $\xi^a_{\rm RW}$, finally obtaining the same result as in \eqref{physicalvelocity}.

\smallskip
Therefore, after considering the slice of the timelike vector field $\xi^a_{(\beta)}$ \eqref{generaltimelikevector} defined in the hypersurface \eqref{hypersurfaces},
the tensor $\Xi_{ab}=\nabla_{b}\xi_a$ is calculated in order to obtain both the expansion $\theta$ and the shear $\sigma_{ab}$  (the twist $\omega_{ab}$ is zero by construction). By following this procedure, the value of $\text{d}\theta/\text{d}\tau$ in the hypersurfaces $f=t-\vec{\epsilon} \cdot \vec{x}$ will be determined. This will provide us with information about the focusing or divergence of the congruence generated by the vector field slice in terms of the scale factor $a(t)$, its derivatives
and the spatial coordinate velocity $\beta$.

\smallskip
However, in the following we shall focus on the mean curvature ${\cal M}_{\xi^a}$. This is due to the fact that although $\sigma_{ab}$ and $\theta$ may not be zero, their contribution to the RE is always non-positive. Moreover, as stressed in Section \ref{raychaudhuriequation}, the implications of the mean curvature ${\cal M}_{\xi^a}$ %, i.e.,\ $-R_{ab}\xi^a\xi^b$,
are more general than those of the other terms. Furthermore, ${\cal M}_{\xi^a}$ is the unique term that may have a positive contribution to the RE \eqref{ray} since for the congruence given rise by the vector field slice \eqref{generaltimelikevector} we have $\omega_{ab}\omega^{ab}=0$.
%Then, if ${\cal M}_{\xi^a} < 0$ the contribution of the mean curvature will be negative \eqref{negativeM} implying that nearby geodesics will have a tendency to focus, or in other words that nearby geodesics suffer an accelerated contraction.
Hence, let us proceed with the calculation of ${\cal M}_{\xi^a_{(\beta)}}$ in order to extract information
about the focusing of these one-parameter dependent congruences.

%%%%%%%%%%%%%%%%%%%%%%%%%%%%%%%%%%%%%%%
%%%%%%%%%%%%%%%%%%%%%%%%%%%%%%%%%%%%%%%

\subsection{Evaluation of the condition ${\cal M}_{\xi^a_{(\beta)}} > 0$}

%%%%%%%%%%%%%%%%%%%%%%%%%%%%%%%%%%%%%%%
%%%%%%%%%%%%%%%%%%%%%%%%%%%%%%%%%%%%%%%
%In order to calculate ${\cal M}_{\xi^a_{(\beta)}}=-R_{ab}\xi^a_{(\beta)}\xi^b_{(\beta)}$ it is only needed a direction and not the behavior of the vector field locally nor in a hypersurface orthogonal to it.
In the following we will obtain a condition on $a(t)$ and the spatial coordinate velocity $\beta$ that guarantees ${\cal M}_{\xi^a_{(\beta)}} > 0$ , i.e., a positive contribution to the RE \eqref{positiveM}.
The motivation of choosing a positive contribution lies in the fact that the fundamental congruence satisfies % and not a negative one is because for the fundamental congruence we have
${\cal M}_{\xi^a_{\text{RW}}}> 0$ in the present epoch as follows from \eqref{fundamentalmean}.

\smallskip

\begin{figure}
    \begin{center}
    \resizebox{8.5cm}{8.5cm}
        {\includegraphics{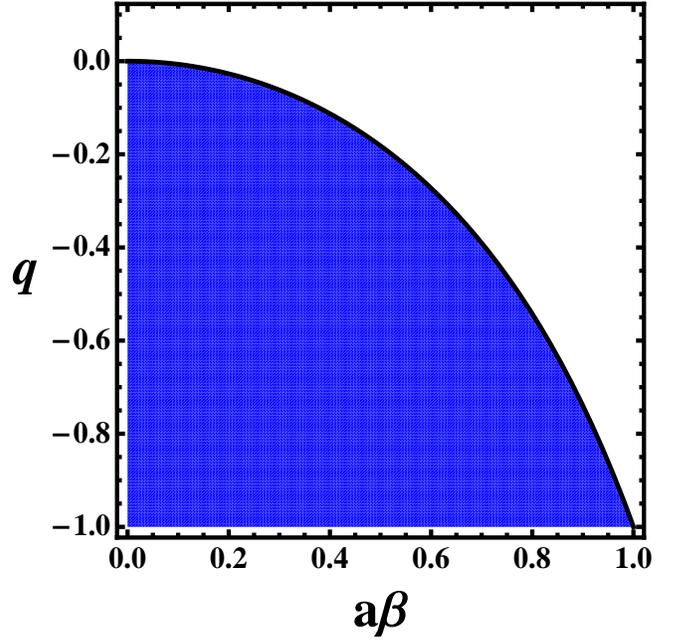}}
    \caption {\footnotesize
    Mean curvature sign for congruence of geodesics generated by \eqref{generaltimelikevector}:
 The line represents the limiting case $q=-2a^2\beta^2/(3-a^2\beta^2)$. For a given spatial coordinate velocity $\beta$ and scale factor $a(t)$ if the deceleration parameter $q$ is below the line (region in blue) a congruence of geodesics with a physical velocity $a\beta$ acquires a positive contribution to the RE from the mean curvature ${\cal M}_{\xi^a}$ or equivalently from the term $-R_{ab}\xi^a\xi^b$. In the region above the line (white color) we depict the opposite case, i.e., negative contribution.
    }
    \label{fig:qbeta}
    \end{center}
\end{figure}

Then, by using the Ricci tensor components derived from the metric \eqref{FLRW} and the expression for the timelike vector field slice given in \eqref{generaltimelikevector}, we get
\begin{eqnarray}
%\begin{aligned}
R_{ab}\xi^a_{(\beta)}\xi^b_{(\beta)}=%\tilde{\gamma}^2 \left(R_{tt}+ \beta^2 R_{xx}\right)\\
%&=
\tilde{\gamma}^2 \left[-3\frac{\ddot{a}}{a}+\beta^2\left(a \ddot{a}+2\dot{a}^2\right)\right]\,.
%\end{aligned}
\end{eqnarray}
Thus, imposing ${\cal M}_{\xi^a_{(\beta)}}=-R_{ab}\xi^a_{(\beta)}\xi^b_{(\beta)} > 0$, yields
\begin{eqnarray}
-3 \frac{\ddot{a}}{a}+\beta^2\left(a \ddot{a}+2\dot{a}^2\right)<0\ \ \Rightarrow\ \ \ddot{a}>\frac{2 a\beta^2 \dot{a}^2 }{3-a^2\beta^2}\,,
\label{decoupled}
\end{eqnarray}
where $a^2\beta^2-3 < 0$ (because of $a\beta<1$) have been used.
This last inequality tells us that the condition $\ddot{a}>0$
is not sufficient for a timelike vector of the form \eqref{generaltimelikevector} to
get a positive contribution to the RE from the space-time geometry ${\cal M}_{\xi^a_{(\beta)}}$.
Consequently, $\ddot{a}$ must be larger than a certain positive value. Therefore, a positive contribution to the RE from space-time geometry \eqref{positiveM} for all timelike congruences is not equivalent to $\ddot{a} > 0$.
% \smallskip
The inequality \eqref{decoupled} can be recast as
\begin{eqnarray}
q<\frac{-2a^2\beta^2}{3-a^2\beta^2}\,,
\label{qI.5}
\end{eqnarray}
%
%where $q$ is the deceleration parameter defined in \eqref{deceleratingparameter}.
which depends only on the physical velocity $a\beta$.
In Figure \ref{fig:qbeta}, the condition \eqref{qI.5} has been depicted. In that figure, the values $(a\beta,q)$ ensuring a positive contribution to the RE from the mean curvature are plotted. %is obtained \eqref{positiveM}.
%
%
% FIGURE 1

Moreover, it is worthwhile to stress that the requirement ${\cal M}_{\xi^a_{(\beta)}}>0$ implies lower values of the deceleration parameter $q$ for timelike vectors approaching the null cone. Then, in the limit $a\beta\rightarrow 1$, i.e.,\ timelike vectors near the null cone, we get $q<-1$. Therefore, provided that $q<-1$ all the congruences of timelike geodesics will have a positive contribution to the RE from space-time geometry which means ${\cal M}_{\xi^a_{(\beta)}} > 0$ for all timelike vectors $\xi^a_{(\beta)}$.
The implication of this condition must not be confused with \eqref{deceleratingparameter} : the condition \eqref{deceleratingparameter}  implied $\text{d}\theta_{\text{RW}}/\text{d}t>0$ for the fundamental congruence whereas in this case the condition $q<-1$ implies
a positive contribution from the space-time geometry, namely ${\cal M}_{\xi^a_{(\beta)}} > 0$, for every congruence of timelike geodesics.
Since the present value of the deceleration parameter is not smaller than $-1$, it remains possible to get ${\cal M}_{\xi^a_{(\beta)}} < 0$ for timelike congruences whose tangent vector field is sufficiently inclined with respect to $\xi^a_{\text{RW}}=\partial/\partial t$. Therefore, using \eqref{qI.5}, %and for given values of the deceleration parameter $q$, the vector field parameter $\beta$ and the scale factor $a(t)$
the following possibilities for the physical velocity $a\beta$ may be fulfilled
\begin{eqnarray}
a\beta<\sqrt{\frac{-3q}{2-q}}\ \ \ \ \ \text{Positive contribution},\,
\label{abI}
\end{eqnarray}
\begin{eqnarray}
a\beta>\sqrt{\frac{-3q}{2-q}}\ \ \ \ \ \text{Negative contribution},
\label{abII}
\end{eqnarray}
\begin{eqnarray}
a\beta=\sqrt{\frac{-3q}{2-q}}\ \ \ \ \ \ \ \ \ \ \text{Zero contribution};
\label{abzero}
\end{eqnarray}
to the RE from the space-time geometry.

%%%%%%%%%%%%eliminado por Franco/////////////////////
%QUITARIA LAS LINEAS ENTRE BARRAS, SOLO CONFUNDEN///// Thus, it may happen that for a test particle following a congruence of geodesics with a tangent vector field whose inclination $\beta$ holds $\beta>\sqrt{-3q/(2-q)}/a$, %the particle moves fast enough with respect to the preferred coordinate system of Robertson-Walker geometry. Hence,
%its contribution to the RE from the mean curvature ${\cal M}_{\xi^a_{(\beta)}}$ will be positive
%and consequently the congruence will tend to focus./////////

If $\beta=0$ in \eqref{decoupled} we obtain $\ddot{a}>0$ which means that the fundamental congruence will experience a positive contribution from the space-time if $\ddot{a}>0$ and therefore in agreement with \eqref{fundamentalmean}.

% FIGURE 2
\begin{figure}
\begin{center}
\resizebox{8.5cm}{8.5cm}
    {\includegraphics{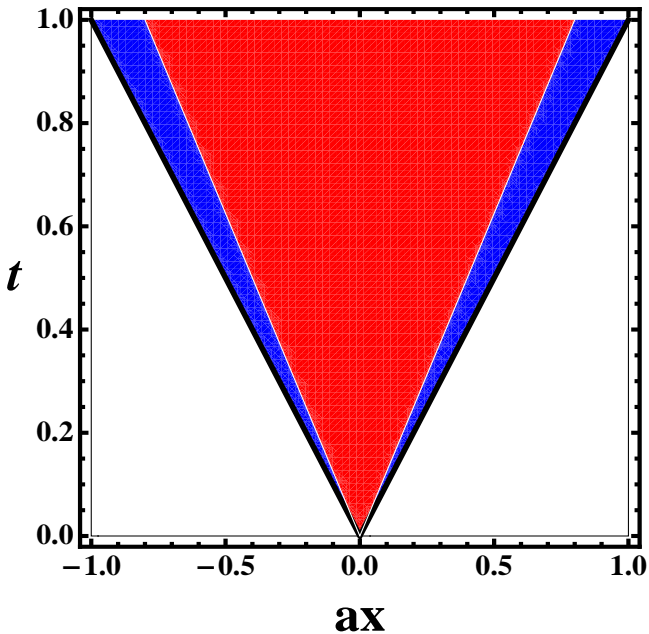}}
    \caption {\footnotesize
   Modifed future light cone in Robertson-Walker metric: The line  $t=ax$ corresponds to the usual light-cone. A positive contribution to the RE (i.e., ${\cal M}_{\xi^a}>0$), for timelike vectors $\xi^a$ lies in the inner (red) area and expression  \eqref{abI} is there satisfied.  For vectors in the outer (blue) zone ${\cal M}_{\xi^a}<0$ and condition \eqref{abII} is now obeyed. The value of the slope of the straight line that separates the two regions corresponds to $1/(a\beta)=1/0.8$. This value is obtained using $q_{0} \simeq -0.6$ in \eqref{abzero} according to WMAP7 \cite{WMAP7}. For smaller values of $q$, i.e. closer to $-1$, the inner region (red) would increase its area and the outer region (blue) would become smaller. For the value $q=-1$, the blue region vanishes and there will be only red zone which corresponds, according to the condition $q<-1$ to all timelike directions having a positive contribution to the RE from the space-time geometry.
    }
\label{fig:nullcone}
\end{center}
\end{figure}

\smallskip
In Figure \ref{fig:nullcone}, a future light cone is represented. Using the value for the present deceleration parameter ($q_{0} \simeq -0.6$), the limiting value
obtained from \eqref{abzero}, $a\beta \simeq 0.8$ separates the regions of different contribution to the RE.
The requirement $a\beta > 0.8$ constitutes a necessary condition to ensure ${\cal M}_{\xi^a} < 0$. However,
due to the non-positive character of all the other terms of the RE for a congruence without vorticity, $\text{d}\theta/\text{d}\tau$
might be negative despite a positive contribution ${\cal M}_{\xi^a} > 0$.

\smallskip
Let us recall that a value of the deceleration parameter $q$ equals to $-1$ is obtained for $a(t) \propto e^{H_{0}t}$ which is the solution in Robertson-Walker cosmological model of the Einstein's equations with cosmological constant $\Lambda=3H_{0}^2$ in vacuum. This solution is usually referred as the de Sitter expanding universe.
This means that a value of $q$ lower than $-1$ implies an expansion rate higher than the rate of expansion of a de Sitter
solution.
Therefore, it seems very unlikely to have $q < -1$ in any epoch of the evolution of the Universe. Let us recall that even in the inflation phase the rate of expansion is expected to be exponential and therefore $q_{\text{inflation}}=-1$.
Thus, let us conclude this section by remarking that
for timelike vectors sufficiently close to the light cone one obtains ${\cal M}_{\xi^a} < 0$ in any epoch of the evolution of the Universe.

%%%%%%%%%%%%%%%%%%%%%%%%%%%%%%%%%%%%%%%%%%%%%%%%%%%%%%%%%%%%%%%%%%%%%%%%%%%%%%
%%%%%%%%%%%%%%%%%%%%%%%%%%%%%%%%%%%%%%%%%%%%%%%%%%%%%%%%%%%%%%%%%%%%%%%%%%%%%%

\section{Conclusions}
\label{conclusions}
%%%%%%%%%%%%%%%%%%%%%%%%%%%%%%%%%%%%%%%%%%%%%%%%%%%%%%%%%%%%%%%%%%%%%%%%%%%%%%%
%%%%%%%%%%%%%%%%%%%%%%%%%%%%%%%%%%%%%%%%%%%%%%%%%%%%%%%%%%%%%%%%%%%%%%%%%%%%%%%
%%%%%%%%%%%%%%%%%%%%%%%%%%%%%%%%%%%%%%%%%%%%%%%%%%%%%%%%%%%%%%%%%%%%%%%%%%%%%%%
In this investigation we have studied the RE and its implications on the evolution of geodesics. We addressed the kinematics of a congruence of timelike geodesics without vorticity in a Robertson-Walker cosmological model with no underlying gravitational theory assumed a priori. From a general point of view, we have established the difference between the implications of a cosmic acceleration $\ddot{a}>0$ and a positive contribution to the RE from the space-time geometry, i.e.\ ${\cal M}_{\xi^a}=-R_{ab}\xi^a\xi^b>0$.
\smallskip

We have then considered the fundamental congruence generated by the vector field $\xi_{\rm RW}^a$. We proved that the tendency to focus or diverge for the fundamental congruence depends of the fact that $q$ is either bigger or smaller than $-1$ respectively.
Given the fact that an approximate value for the deceleration parameter today $q_{0} \simeq -0.6$ can be found either by assuming GR or just by using Supernovae Ia data in theory-independent calculations, one concludes that although
the fundamental congruence is expanding because of $\theta_{\rm RW}=3 \dot{a}/a > 0$, it has a tendency to focus since $\text{d}\theta_{\rm RW}/\text{d}t<0$.

\smallskip
Moreover, we have examined the congruence of timelike geodesics generated by a vector field orthogonal to a one-parameter $\epsilon$ dependent spacelike hypersurface. The spatial coordinate velocity $\beta=\epsilon/a^{2}(t)$ represented -- at each point of the hypersurface -- the inclination of the vector field with respect to the preferred direction of Robertson-Walker models $\xi^a_{\rm RW}$.
In the case of the slice of a vector field to be normalized and orthogonal to the hypersurface,
it is possible to determine the condition for the deceleration parameter $q$ which ensures
the contribution of the space-time geometry to the RE, ${\cal M}_{\xi^a}=-R_{ab}\xi^a\xi^b$ to be positive.
% , although the sign of $\text{d}\theta/\text{d}\tau$ remains unknown it was possible
%
The aforementioned condition depends both upon the spatial coordinate velocity $\beta$ and the scale factor $a(t)$ but
only through the physical spatial velocity $a\beta$. In congruences without vorticity, if ${\cal M}_{\xi^a} > 0$ does not hold all the contributions to the RE are non-positive since $\omega_{ab}=0$ and the remaining terms are non-positive as we discussed.
We found that this condition is not satisfied in the present Universe provided that $a\beta>0.8$. This means that congruences of geodesics without vorticity with its tangent vector fulfilling $a\beta>0.8$ will not tend to diverge. In the Robertson-Walker preferred framework this situation occurs for
geodesics moving {\it near} the light cones. These congruences will tend to focus provided that the other terms of the RE are negative. Besides, a negative contribution from space-time geometry to the RE would mean that the attractive character of gravity is restored at high speeds for  $-1<q<0$. The term high speeds must be understood with respect to the preferred framework of a Robertson-Walker cosmological model which is usually identified with the CMB because of its high isotropy.
\smallskip

In the previous sense, systems may decouple from the cosmological accelerated expansion. This result can help to understand the structure formation of system moving fast, that is favored with respect to slow systems. Even taking into account the present value of the deceleration parameter $q_{0}$, $a\beta > 0.8$ gives a negative contribution to the RE from space-time geometry. This means that in order to get a negative contribution to RE from the space-time geometry, it is necessary to move with a velocity bigger than $0.8\,c$ with respect to the preferred framework of Robertson-Walker space-times.
Although this value is quite large,  the deceleration parameter $q$ was smaller in the past and consequently the velocity respect the CMB necessary to accomplish a negative contribution to the RE from space-time geometry was also smaller.

To conclude, we would like to highlight the fact that the results obtained in this investigation provide us with a mechanism for a gravitational system
to decouple from the accelerated expansion of the Universe. This mechanism does not need to appeal to the
gravitational interaction of the system itself. Although the decoupling of astrophysical systems from the cosmic acceleration
is usually carried out by the stronger effect of their own gravitational forces, our investigation may help to understand how geodesic motion in an
expanding Universe influences this decoupling.

%%%%%%%%%%%%%%%%%%%%%%%%%%%%%%%%%%%%%%%%%%%%%%%%%%%%%%%%%%%%%%%%%%%%%%%%%%%%%%%%
%%%%%%%%%%%%%%%%%%%%%%%%%%%%%%%%%%%%%%%%%%%%%%%%%%%%%%%%%%%%%%%%%%%%%%%%%%%%%%%
\vspace{0.75cm}
%\begin{acknowledgments}
{\bf Acknowledgments.}
We would like to thank prof. G.\ F.\ R.\ Ellis for drawing our attention to reference \cite{Ellis-Elst}.
This work has been supported by MICINN (Spain) projects numbers
FIS2011-23000, FPA2011-27853-C02-01 and Consolider-Ingenio MULTIDARK CSD2009-00064.
FDA would like to thank Theoretical Physics I department, Complutense University of Madrid for providing all the required academic resources.
AdlCD acknowledges financial support from NRF and URC research fellowships (South Africa).
JARC acknowledges the kind hospitality of ACGC/University of Cape Town while elaborating the manuscript.

%%%%%%%%%%%%%%%%%%%%%%%%%%%%%%%%%%%%%%%%%%%%%%%%%%%%%%%%%%%%%%%%%%%%%%%%%%%%%%%%%%%%%%%
%%%%%%%%%%%%%%%%%%%%%%%%%%%%%%%%%%%%%%%%%%%%%%%%%%%%%%%%%%%%%%%%%%%%%%%%%%%%%%%%%%%%%%%
%%%%%%%%%%%%%%%%%%%%%%%%%%%%%%%%%%%%%%%%%%%%%%%%%%%%%%%%%%%%%%%%%%%%%%%%%%%%%%%%%%%%%%%

\end{document}